\documentclass[12pt]{article}
\usepackage{indentfirst} 
\usepackage{amsfonts,amssymb,amsmath,amsthm}
\usepackage{bm}          
\usepackage{cite}
\usepackage{url}
\usepackage{enumitem}
\usepackage[usenames,dvipsnames]{xcolor}
\usepackage{todonotes}
\usepackage{multicol}
\usepackage{MnSymbol}
\usepackage{hyperref}
\usepackage{textcomp}
\usepackage[titletoc,title]{appendix}
\usepackage{smartref}               \addtoreflist{section}   

\theoremstyle{definition}
\newtheorem*{propertiess}{Properties}

\newcommand{\refconchap}[1]{\hyperref[#1]{\textcolor{\corlink}{\sectionref{#1}.}\ref*{#1}}}   

\newcommand{\margenizquierda}{2.8cm}   
\newcommand{\margenderecha}{2.8cm}       
\newcommand{\margensuperior}{3cm}      
\newcommand{\margeninferior}{3.5cm}    
\usepackage{geometry} 				

\allowdisplaybreaks

\makeatletter
\newcommand{\raisemath}[1]{\mathpalette{\raisem@th{#1}}}
\newcommand{\raisem@th}[3]{\raisebox{#1}{$#2#3$}}
\makeatother

 \def\QED{{\boldmath$\rule{0.5em}{0.5em}$}}                                
 \def\markatright#1{\leavevmode\unskip\nobreak\quad\hspace*{\fill}{#1}}    
 \def\qed{\markatright{\QED}}                                              

\newcommand{\R}{\mathbb{R}}     

\newcommand{\corurl}{blue}
\newcommand{\corcite}{ForestGreen}
\newcommand{\corlink}{blue}

\def\ed{\bm{{\rm{d}}} }

\hypersetup{linktocpage,colorlinks,urlcolor=\corurl,citecolor=\corcite,linkcolor=\corlink,
	pdftitle={Geometric Dirac},
	pdfauthor={J.F. Barbero, B. Díaz, J. Margalef-Bentabol, E.J.S. Villaseñor}}

\usepackage{titling}
\usepackage{authblk}
\title{Dirac's algorithm in the presence of boundaries: a practical guide to a geometric approach}
\author[1,4]{J. Fernando Barbero G.}
\author[1]{Bogar D\'{\i}az}
\author[2,4,5]{Juan Margalef-Bentabol}
\author[3,4]{Eduardo J.S. Villase\~nor\,}
\affil[1]{Instituto de Estructura de la Materia, CSIC. Serrano 123, 28006 Madrid, Spain}
\affil[2]{Laboratory of Geometry and Dynamical Systems, Department of Mathematics, EPSEB, Universitat Polit\`ecnica de Catalunya, BGSMath, Barcelona, Spain}
\affil[3]{Universidad Carlos III de Madrid. Avda.\  de la Universidad 30, 28911 Legan\'es, Spain}
\affil[4]{Grupo de Teor\'{\i}as de Campos y F\'{\i}sica Estad\'{\i}stica. Instituto Gregorio Mill\'an (UC3M). Unidad Asociada al Instituto de Estructura de la Materia, CSIC, Madrid, Spain}
\affil[5]{Institute for Gravitation and the Cosmos \& Physics Department, Penn State, University Park, PA 16802, USA}

\date{}                     
\date{September 4, 2019}

\begin{document}

\maketitle

\begin{abstract}
The goal of this paper is to propose and discuss a practical way to implement the Dirac algorithm for constrained field models defined on spatial regions \emph{with boundaries}. Our method is inspired in the geometric viewpoint developed by Gotay, Nester, and Hinds (GNH) to deal with singular Hamiltonian systems.  We pay special attention to the specific issues raised by the presence of boundaries and provide a number of significant examples---among them field theories related to general relativity---to illustrate the main features of our approach.
\end{abstract}

\section{Introduction}

Constrained field theories are very important examples of singular dynamical systems, i.e. those for which the accelerations cannot be uniquely written in terms of the generalized coordinates and velocities. In the guise of gauge theories, they are an essential ingredient of the standard model of fundamental interactions and connection formulations of general relativity. They also play a central role in many other contexts such as condensed matter physics. An interesting approach to their study, and a possible starting point for quantization, is to present their dynamics in Hamiltonian form.

A satisfactory understanding of the Hamiltonian description of singular systems was achieved by Dirac \cite{Dirac}, who culminated the work of other authors (in particular Anderson and Bergmann \cite{AndersonBergmann}) and introduced his celebrated ``algorithm'' to systematically deal with such models. Although, arguably, the logic of Dirac's method is clear and the geometric meaning of its main ingredients has been understood for quite a long time, several issues crop up in practice, in particular if spatial boundaries are present.

The original approach developed by Dirac deals with singular \emph{mechanical} models with finite dimensional configuration spaces. Although in his own words\footnote{This is actually the attitude that, for often reasonable and justifiable practical purposes, many authors take, see for instance \cite[p.494]{Pons1}.} \cite[p.6]{Dirac}
\begin{center}
\begin{minipage}{0.93\linewidth}\emph{It is then merely a formal matter to pass from this finite number of degrees of freedom to the infinite number of degrees of freedom which we need for a field theory}
\end{minipage}
\end{center}
\noindent in practice, there are subtleties that must be addressed. Let us mention some of them\footnote{Multisymplectic methods \cite{GotayMarsden} are also relevant in this context, however, the quantization of field theories still relies heavily on the Hamiltonian formulation and the, somehow related, path integrals.}.

The first is the fact that configuration spaces for field theories are infinite dimensional manifolds. A crucial difference between them and their much simpler finite dimensional counterparts is that the former are modeled on linear functional spaces---often Banach spaces with norms that must be explicitly specified---instead of finite dimensional vector spaces, where all the norms are equivalent. Although we will gloss over such functional analytic issues, it should be kept in mind that they must be carefully taken into account if a rigorous and complete description is sought (the interested reader is referred to \cite{BPV1} for a detailed discussion of some relevant examples solved with the help of the GNH method \cite{Gotay1,Gotay2,Gotay3,Gotay4}). Our purpose here is to sidestep these fine mathematical points while providing an easy-to-use procedure to obtain all the other elements of the phase space formulation for field theories with boundaries: constraints, Hamiltonians, and Hamiltonian vector fields.

Another source of difficulties in this context is the need to deal with the boundary conditions necessary to completely specify physical configurations and dynamics. These may be incorporated in the definition of the configuration space or may appear when looking for the stationary points of the action. In both cases, one has to check the consistency of the dynamics defined by the field equations \emph{and} the boundary conditions. A famous instance of one such incompatibility was detected (and solved) by Regge and Teitelboim in the classical paper \cite{reggeteitelboim} where they showed the need to add a suitable surface term to the Hamiltonian of asymptotically flat general relativity in order to have dynamics consistent with Einstein's equations. In this regard, it is important to understand the role played by the concept of \emph{functional differentiability} introduced and used by those authors. A related issue is the interpretation of the boundary conditions in the Hamiltonian framework: are they constraints? If the answer is in the affirmative, are they first or second class? Also, does their dynamical stability give rise to additional secondary constraints? The correct answer to these questions is important if the final goal of the Hamiltonian analysis is to use it as a step towards quantization.

When dealing with bounded systems, it is often interesting to find out to what extent it is possible to locate physical degrees of freedom either at the bulk or at the boundaries\cite{cuerdamasas,tesisJuan}. In this regard, it is appropriate to mention condensed matter physics (topological insulators and related systems). From the point of view of the reduced phase space the possibility of assigning physical degrees of freedom either to the bulk or the boundary hinges on having an easy-to-interpret reduced phase space such as the cotangent bundle of a product of manifolds (possibly infinite dimensional) associated in a natural and unambiguous way with the bulk and the boundaries. In order to consider this issue, it is necessary to develop a reliable way to get the Hamiltonian description of field theories in bounded regions.

An interesting approach to the study of the Hamiltonian formulation of singular Lagrangian systems (and more general instances of singular dynamical models) is the one developed by Gotay, Nester, and Hinds \cite{Gotay1,Gotay2,Gotay3}. Their method relies heavily on the use of geometric concepts and is carefully crafted to deal with field theories. Its central idea is to look directly for a Hamiltonian vector field---living on a submanifold of a presymplectic manifold---whose integral curves, appropriately projected, give the solutions to the equations of motion. In this respect, the GNH method is somewhat far from the philosophy of the Dirac approach which is to provide an avenue to the quantization of singular systems based on the use of quantum versions of the constraints defined on the \emph{full} phase space. We will use the main ideas of \cite{Gotay1} and adapt them to the context of the Dirac method.

\medskip

\noindent The main observations that we make in this paper are:

\begin{itemize}
\item The Dirac method can be carried out in a way that essentially mirrors the steps of the GNH algorithm and takes advantage of its most useful features, in particular its geometric perspective.
\item There is no need to explicitly use Poisson brackets to obtain the basic elements of the Hamiltonian description of a mechanical system or field theory (constraints, Hamiltonian vector fields, and Hamiltonian). In the presence of boundaries this helps to avoid problems associated with the use of formal expressions, in terms of functional derivatives, for the symplectic form.
\item The method that we propose here is also slick and user friendly---in practice much simpler than the standard approach---for the usual mechanical systems and field theories (and not only when boundaries are present).
\end{itemize}

We will illustrate the above by discussing a number of examples from our perspective and compare our results with the existing literature when available.

The plan of the paper is the following. After this introduction, we start in section \ref{Sect_General} by spelling out the method that we use with the help of an example: a physical pendulum described as a constrained system. Section \ref{Sect_Dirichlet} will be devoted to study the scalar field with Dirichlet boundary conditions and show how our approach avoids the use of awkward and, sometimes, \emph{ad hoc} procedures to deal with the tower of boundary conditions characteristic of this model. Sections \ref{sec_CS} and \ref{sec_Pontryagin} are devoted to a discussion of an interesting problem that has been recently considered in the literature \cite{Corichi}: the detailed study of the relationship between the Hamiltonian formulations of the dynamics defined by the Pontryagin action in a four-dimensional manifold with boundaries and the Chern-Simons description of the same model. The main difference between our treatment and the one presented in \cite{Corichi} is that we will not need to invoke Regge-Teitelboim differentiability. We devote Section \ref{Sect_Palatini} to our final example: 2+1 gravity in terms of triad-connection variables with boundary conditions. We end with our final comments and conclusions. The notation used throughout the paper is essentially standard, but carefully tries to avoid the use of symbols that may carry with them unwanted meanings.

\section{From the GNH approach to the Dirac method}\label{Sect_General}{

In this section, we reinterpret the Dirac approach to the treatment of singular dynamical systems from the perspective provided by the GNH method. We partially follow \cite[ch.V]{Gotay1}; other relevant details can be found in \cite{Gotay2,Gotay3}. Our presentation is tailored to facilitate the implementation of the Dirac algorithm for field theories in the presence of boundaries (see \cite{BPV1,BMV1} for detailed discussions of the GNH procedure for this purpose).

The geometric arena to describe a Lagrangian mechanical system is the tangent bundle $T\mathcal{Q}$ of a differentiable manifold $\mathcal{Q}$  whose points represent physical configurations. The Lagrangian $L$ of a time-independent model is a real function $L:T\mathcal{Q}\rightarrow\mathbb{R}$. 
 Given a space $\mathcal{P}_{q_1,q_2}$ of sufficiently smooth curves in $\mathcal{Q}$ connecting two fixed points $q_1,q_2\in \mathcal{Q}$, we define the action associated with a Lagrangian $L$ as
\[
S:\mathcal{P}_{q_1,q_2}\rightarrow\mathbb{R}:c\mapsto\int_{t_1}^{t_2}L\big(c(t),\dot{c}(t)\big)\mathrm{d}t\,.
\]
The dynamical evolution of the system from $q_1$ at time $t_1$ to $q_2$ at time $t_2$ is given by curves corresponding to stationary points of $S$. The familiar Euler-Lagrange equations are necessary conditions for $S$ to be stationary.

Regular Lagrangians are those for which the Euler-Lagrange equations allow us to solve for the accelerations $\ddot{q}(t)$ in terms of the configuration variables and velocities, whereas singular systems are those for which this is not possible. In this case, it often happens that some of the equations of motion are relations involving only positions and velocities whereas, in other instances, not all the accelerations are independent.

The Hamiltonian formalism is formulated in \emph{phase space}, the cotangent bundle $T^*\mathcal{Q}$. The first step to obtain it is to define the momenta by using what in the mathematical parlance is known as the \emph{fiber derivative} $F\!L$ associated with the Lagrangian $L$. This is a map $F\!L:T\mathcal{Q}\rightarrow T^*\mathcal{Q}$ where $p:=F\!L_q(v)\in T_q^*\mathcal{Q}$ is defined by
\[
\langle p|w\rangle:=\frac{\mathrm{d}}{\mathrm{d}t}L(q,v+tw)\Big|_{t=0}\,,\quad v,w\in T_q\mathcal{Q}\,.
\]
In the following, we will drop the $q$ subindex and write $F\!L$. Whenever this map is a diffeomorphism we can define the Hamiltonian, a real function on phase space, as $H=E\circ F\!L^{-1}$ where $E$ is the energy (conserved under the evolution defined by the Euler-Lagrange equations) given by
\[
E:T\mathcal{Q}\rightarrow \mathbb{R}:(q,v)\mapsto \langle F\!L(v)|v\rangle-L(q,v)\,.
\]
Singular systems can also be characterized as those for which the fiber derivative fails to be invertible\footnote{This may happen for a number of reasons and in different ways. For instance, the fiber derivative may not be defined at some points of $T\mathcal{Q}$, fail to be injective, be injective but not onto\ldots}. In this case the relation
\begin{equation}\label{definitionHamiltonian}
  H\circ F\!L=E
\end{equation}
defines the Hamiltonian only on the image of $FL$. For the sake of clarity, we will suppose in the following that $FL$ defines a smooth submanifold $\mathcal{M}_0:=F\!L(T\mathcal{Q})$ of $T^*\mathcal{Q}$. This submanifold is represented in the Dirac approach as the null set of some functions $\phi_n$ known as \emph{primary constraint} functions. It is obvious that, in this case, equation\ \eqref{definitionHamiltonian} does not define a Hamiltonian on the full $T^*\mathcal{Q}$ but only on $\mathcal{M}_0$.

When implemented in the context of the Hamiltonian dynamics in the contangent bundle $T^*\mathcal{Q}$, the GNH method takes $\mathcal{M}_0$, the  pullback $\omega$ of the canonical symplectic form $\Omega$ to  $\mathcal{M}_0\,$, and the Hamiltonian $H$ (uniquely defined on $\mathcal{M}_0$) as the starting point \cite{Gotay4}. Its aim is to find a (maximal) submanifold $\mathcal{M}$ of $\mathcal{M}_0$ and a vector field $X$ on $\mathcal{M}_0$ [i.e.\ $X\in\mathfrak{X}(\mathcal{M}_0)$] such that the following equation holds\footnote{Here and in the following we will denote the exterior derivative in phase space as \bm{\mathrm{d}} and the inner derivative as ${\bm{\imath}}$.}:
\begin{equation}\label{ecGNH}
(\bm{\imath}_X\omega-\bm{\mathrm{d}}H)|\mathcal{M}=0\,,
\end{equation}
with $X$ \emph{tangent to} $\mathcal{M}$. The integral curves of $X$ corresponding to initial data on $\mathcal{M}$ give the dynamical evolution of the system once projected onto $\mathcal{Q}$.

It is possible to adapt some of the features of the GNH method---in particular its geometric flavor and the idea of viewing dynamical consistency as a tangency condition---to the implementation of the Dirac algorithm \cite{Munhoz,Gotay1}.
The steps to do this are the following: Once the primary constraint submanifold $\mathcal{M}_0$ is identified, one looks for a function $H$ \emph{defined on the whole phase space} and satisfying condition \eqref{definitionHamiltonian}. This function will play the role of the Hamiltonian. Notice that it is uniquely defined \emph{only on} $\mathcal{M}_0$.

For regular Lagrangians it is possible to get the dynamics in Hamiltonian form by looking for the stationary points of the phase space action
\begin{equation}\label{Ham_action}
S_\Gamma(q,p):=\int_{t_1}^{t_2}\Big(\langle p(t)|\dot{q}(t)\rangle-H\big(q(t),p(t)\big)\Big)\mathrm{d}t
\end{equation}
defined in a space $\mathcal{P}_\Gamma$ of sufficiently smooth curves on $T^*\mathcal{Q}$. Whenever primary constraints are present the dynamics must be such that they are preserved by the evolution, i.e.\ for all $t\in[t_1,t_2]$ we must have
\[
\phi_n\big(q(t),p(t)\big)=0\,.
\]
The standard way to look for stationary points of a real function subject to constraints is to enforce them with the help of suitable Lagrange multipliers. As the action \eqref{Ham_action} is defined \emph{in a space of curves} these multipliers  $u_n(t)$ must be functions of time. According to this, the action principle that we must use to get the Hamiltonian dynamics of a singular system (see, for instance, \cite{Batlle}) is
\begin{equation}\label{Ham_action_sing}
\widetilde{S}_\Gamma(q,p,u):=\int_{t_1}^{t_2}\Big(\langle p(t)|\dot{q}(t)\rangle-H\big(q(t),p(t)\big)-u_n(t)\phi_n\big(q(t),p(t)\big)\Big)\mathrm{d}t\,,
\end{equation}
leading to the following equations of motion
\begin{subequations}\label{equations_u}
\begin{align}
& \dot{q}=\frac{\partial H}{\partial p}+u_n\frac{\partial \phi_n}{\partial p}\,,\label{equations_u1}\\
& \dot{p}=-\frac{\partial H}{\partial q}-u_n\frac{\partial \phi_n}{\partial q}\,,\label{equations_u2}\\
&\phi_n\big(q(t),p(t)\big)=0\,.\label{equations_u3}
\end{align}
\end{subequations}
In the context of the implementation of the Dirac algorithm, we will refer to the $u_n(t)$ as \emph{Dirac multipliers}. They must not be confused with other variables that may appear in the Lagrangian\footnote{Although the present discussion is to a certain extent trivial, in the perception of the authors there is some confusion in the literature regarding the meaning of the $u_n(t)$. Even in \cite{Dirac} the $u_n(t)$ are presented in a somewhat mysterious way as, for instance, when it is stated that the Poisson bracket of $u_n(t)$ with the $q$'s or the $p$'s is not defined \cite[p.12]{Dirac}.}, in fact,  $u_n(t)$ have nothing to do with the configuration manifold $\mathcal{Q}$.

The resolution of \eqref{equations_u} is not straightforward. In particular, one does not expect these equations to have solutions for arbitrary initial data $(q(t_1),p(t_1))$ \emph{even if they satisfy---as they must---the constraints} \eqref{equations_u3}. Notice also that these equations involve the Dirac multipliers $u_n(t)$, which must be partially or totally fixed to guarantee their consistence. The Dirac algorithm can be viewed as a systematic approach to address these two issues. We present a geometric interpretation thereof in the following. The stationary points of \eqref{Ham_action_sing} can be found by looking for a time-dependent vector field $X\in\mathfrak{X}(T^*\mathcal{Q})$ satisfying
\begin{equation}\label{ecs_X_u}
\bm{\imath}_{X}\Omega-\bm{\mathrm{d}}H-u_n(t) \bm{\mathrm{d}}\phi_n=0\,,
\end{equation}
and finding its integral curves. The consistency of the dynamics implies that $X$ must be tangent to $\mathcal{M}_0$ (in other words, the primary constraints must be preserved under time evolution, i.e.\  $\phi_n(q(t),p(t))=0$ for all $t\in[t_1,t_2]$). In the process of solving \eqref{ecs_X_u} two alternative things may happen:
\begin{enumerate}
  \item Any solution $X$ to \eqref{ecs_X_u} is tangent to $\mathcal{M}_0$ (i.e.\ $\bm{\pounds}_X\phi_n=\bm{\imath}_X\bm{\mathrm{d}}\phi_n=0$).
  \item The tangency condition $\bm{\imath}_X\bm{\mathrm{d}}\phi_n=0$ only holds on a certain proper submanifold $\mathcal{M}_1\subset \mathcal{M}_0$ and/or for some specific values of $u_n(t)$.
\end{enumerate}
In the first case we are done; the integral curves of $X$ describe the evolution of initial data satisfying the constraints $\phi_n=0$. In the second case we must check if the vector field $X$ is tangent, not only to $\mathcal{M}_0$, but also to $\mathcal{M}_1$ (described, in practice, as the null set of \emph{secondary constraints} $\phi_j$). If this is the case we are done, if not, we must iterate the procedure by requiring appropriate tangency conditions until it stops. This will happen when, after requiring that the vector field $X$ is tangent to the submanifold $\mathcal{M}_m$ (i.e.\ $\bm{\imath}_X\bm{\mathrm{d}}\phi_{j_m}=0$ for the secondary constraints defining $\mathcal{M}_m$), it will also be tangent to the submanifold $\mathcal{M}_{m+1}$ determined at this step. Generically some Dirac multipliers will be fixed in this process.

\subsection{An example: the pendulum}

We illustrate now how the GNH idea can be used to implement the Dirac algorithm with a simple mechanical example: a plane pendulum described as a constrained system.\footnote{See \cite{Matschull}. Another interesting mechanical example displaying a genuine gauge symmetry is discussed in \cite{BPV2}.}

Let us consider the configuration space $\mathcal{Q}=\mathbb{R}^3$ (so that $T\mathcal{Q}=\mathbb{R}^3\times \mathbb{R}^3\simeq \mathbb{R}^6$) and take the Lagrangian
\begin{equation}
L:\mathbb{R}^3\times \mathbb{R}^3\rightarrow \mathbb{R}:(q,v)\mapsto \frac{1}{2}m(v_x^2+v_y^2)-m g y+\zeta(x^2+y^2-\ell^2)\,,
\label{lag_pendulum}
\end{equation}
where $q=(x,y,\zeta)$, $v=(v_x,v_y,v_\zeta)$, $m$ and $\ell$ are the mass and length of the pendulum and $g$ is the acceleration of gravity. Here $x$ and $y$ are the cartesian coordinates of the pendulum bob on the plane and $\zeta$ is an auxiliary variable introduced to enforce the condition that the length of the pendulum is $\ell$. In practice, $\zeta$ plays the role of a Lagrange multiplier although it is important to keep in mind that it is a configuration variable on an equal footing with the remaining ones, $x$ and $y$. Its dynamics (as given by the Euler-Lagrange equations) certainly differs qualitatively from that of $x$ and $y$, but this is a consequence of the singular character of the system in this representation, a fact that is only revealed after the equations of motion are obtained. Indeed the Euler-Lagrange equations are
\begin{subequations}\label{eulerpendulum}
\begin{align}
m\ddot{x}-2\zeta x&=0\,,\label{eulerpendulum1}\\
m\ddot{y}-2\zeta y+m g&=0\,,\label{eulerpendulum2}\\
x^2+y^2-\ell^2&=0\,.\label{eulerpendulum3}
\end{align}
\end{subequations}
As we can see the acceleration $\ddot{\zeta}$ does not appear in \eqref{eulerpendulum} hence, the Lagrangian \eqref{lag_pendulum} is singular. Equation \eqref{eulerpendulum3} can be solved by introducing a new real variable $\theta\in[0,2\pi)$ and writing $x=\ell\sin\theta$, $y=-\ell\cos\theta$. The first two equations \eqref{eulerpendulum1},\eqref{eulerpendulum2} are then equivalent to
\begin{subequations}
\begin{align}
&\ddot{\theta}+ \frac{g}{\ell}\sin\theta=0\,,\label{eq.1}\\
&\zeta=-\frac{m}{2}(\dot{\theta}^2+\frac{g}{\ell}\cos\theta)\,.\label{eq.2}
\end{align}
\end{subequations}
The first one describes the motion of a physical pendulum. The second one---expressing $\zeta$ in terms of $\theta$ and $\dot{\theta}$---and the parametrization $x=\ell\sin\theta$, $y=-\ell\cos\theta$ allow us to get from \eqref{eulerpendulum} the tension $\mathbf{T}:=(2x\zeta,2y\zeta)$ of the string or bar connecting the pendulum to its support as a function of $\theta$ and $\dot{\theta}$. In fact, this provides the physical interpretation of $\zeta$ through its relationship with $\mathbf{T}$. From a practical point of view the possibility of finding $\mathbf{T}$ may be a good reason to prefer the Lagrangian \eqref{lag_pendulum} to the usual regular one defined on the circumference $\mathbb{S}^1$, despite the proliferation of configuration variables.

Let us consider now the cotangent bundle $T^*\mathcal{Q}\simeq \mathbb{R}^3\times \mathbb{R}^3$ endowed with the canonical symplectic form
\begin{equation}
\Omega=\bm{\mathrm{d}}x\wedge\bm{\mathrm{d}}p_x+\bm{\mathrm{d}}y\wedge\bm{\mathrm{d}}p_y+\bm{\mathrm{d}}\zeta\wedge\bm{\mathrm{d}}p_\zeta\,.
\label{symplecticpendulum}
\end{equation}
The fiber derivative $F\!L:T\mathcal{Q}\to T^*\mathcal{Q} $ is given by
\begin{equation}\label{FL}
\begin{array}{l}
\left\langle F\!L\left(v\right) | w \right\rangle=m(v_xw_x+v_yw_y)
\end{array}
\end{equation}
where the momentum $\mathbf{p}:=FL(v)$ can be represented by the vector $p=(mv_x,mv_y,0)\in\mathbb{R}^3$, so that $p_x=mv_x$, $p_y=mv_y$ and $p_\zeta=0$. The condition $p_\zeta=0$ is a primary constraint. The primary constraint submanifold $\mathcal{M}_0$ is then given by
\begin{equation}
\mathcal{M}_0= \left\{ \left( q,p \right) \in T^*\mathcal{Q}: \phi_1:=p_\zeta=0 \right\}.
\end{equation}
The Hamiltonian is uniquely defined \emph{only} on $\mathcal{M}_0$. An extension to the full phase space of the Hamiltonian on $\mathcal{M}_0$ is
\begin{equation}\label{HM1}
H:\mathbb{R}^3\times\mathbb{R}^3 \rightarrow\mathbb{R}:((x,y,\zeta),(p_x,p_y,p_\zeta))\mapsto \frac{1}{2m}(p_x^2+p_y^2)+mgy-\zeta(x^2+y^2-\ell^2)\,,
\end{equation}
A simple computation gives
\begin{equation}\label{dH}
\bm{\mathrm{d}}H=-2\zeta x \bm{\mathrm{d}}x +(mg-2\zeta y)\bm{\mathrm{d}}y-(x^2+y^2-\ell^2)\bm{\mathrm{d}}\zeta+\frac{p_x}{m}\bm{\mathrm{d}}p_x+\frac{p_y}{m}\bm{\mathrm{d}}p_y \,.
\end{equation}
A vector field $X\in\mathfrak{X}(T^*\mathcal{Q})$ has the form
\begin{equation}\label{vector_field}
  X=X_x\partial_x+X_y\partial_y+X_\zeta\partial_\zeta+X_{p_x}\partial_{p_x}+X_{p_y}\partial_{p_y}+X_{p_\zeta}\partial_{p_\zeta} \,,
\end{equation}
so that $\displaystyle  \bm{\imath}_X\Omega=-X_{p_x}\bm{\mathrm{d}}x-X_{p_y}\bm{\mathrm{d}}y-X_{p_\zeta}\bm{\mathrm{d}}\zeta+X_x\bm{\mathrm{d}}p_x+X_y\bm{\mathrm{d}}p_y+X_\zeta\bm{\mathrm{d}}p_\zeta\,.$ Equation \eqref{ecs_X_u} becomes now
\begin{align*}
  & -X_{p_x}\ed x-X_{p_y}\ed y-X_{p_\zeta}\ed \zeta+X_x\ed p_x+X_y\ed p_y+X_\zeta\ed p_\zeta \\
  = & -2\zeta x \ed x +(mg-2\zeta y)\ed y-(x^2+y^2-\ell^2)\ed \zeta+\frac{p_x}{m}\ed p_x+\frac{p_y}{m}\ed p_y+u\ed p_\zeta\,,
\end{align*}
where $u$ is a Dirac multiplier introduced to enforce the primary constraint $p_\zeta=0$. By comparing the two sides of the previous expression we immediately get
\begin{align*}
X_x&=\frac{p_x}{m}´\,, &X_y&=\frac{p_y}{m}\,,  &X_\zeta &=u\,,\\
X_{p_x}&=2\zeta x\,,       &X_{p_y}&=2\zeta y-mg\,,     &X_{p_\zeta}&=x^2+y^2-\ell^2\,.
\end{align*}
It is very important to notice now that $X$ is not tangent to most of the points in $\mathcal{M}_0$ but only to those where the \emph{tangency condition} $0=\bm{\imath}_X  \ed \phi_1= \bm{\imath}_X\ed p_\zeta=X_{p_\zeta}$ holds. This immediately gives
\begin{equation}
\mathcal{M}_1:=\mathcal{M}_0\cap\{(q,p)\in T^*\mathcal{Q}: \phi_2:=x^2+y^2-\ell^2=0 \}\,.
\label{M1}
\end{equation}
We thus obtain the secondary constraint $\phi_2=0$. We demand now that $X$ is tangent to $\mathcal{M}_1$, i.e.\ $0=\bm{\imath}_X  \mathrm{d} \phi_2$, this gives
\begin{align*}
0=\bm{\imath}_{X}  \bm{\mathrm{d}} \phi_2 &=2x X_x +2yX_y=\frac{2}{m}(xp_x+yp_y)=:\frac{2}{m}\phi_3\,\
\Rightarrow \mathcal{M}_2:=\mathcal{M}_1\cap\{(q,p)\in T^*\mathcal{Q}: \phi_3=0 \}\,.
\end{align*}
Iterating this process, we find
\begin{align*}
0=\bm{\imath}_{X} \bm{\mathrm{d}} \phi_3 &= x X_{p_x}+X_x p_x+y X_{p_y}+X_y p_y= \frac{p_x^2+p_y^2}{m}-mgy+ 2\zeta \left(\ell^2+\phi_2\right)\,,\\
\Rightarrow\mathcal{M}_3&:=\mathcal{M}_2\cap\{(q,p)\in T^*\mathcal{Q}: \phi_4:=\frac{p_x^2+p_y^2}{m}+2\ell^2\zeta-mgy=0 \}\,,
\end{align*}
and, finally,
\begin{equation*}
0=\bm{\imath}_{X}  \bm{\mathrm{d}} \phi_4=2\frac{p_x X_{p_x}+p_yX_{p_y}}{m}+2\ell^2X_\zeta-mgX_y=\frac{4\zeta}{m} \phi_3+2\ell^2u-3gp_y\,.
\end{equation*}
The last expression fixes the Dirac multiplier to the value $u=3gp_y/2\ell^2$ and gives no more constraints.

The result of the preceding analysis tells us that we will have consistent Hamiltonian dynamics for the pendulum in the phase space submanifold given by
\[\mathcal{C}:=\left\{(q,p)\in T^*\mathcal{Q} : p_\zeta=0, x^2+y^2-\ell^2=0, xp_x+yp_y=0, \frac{p_x^2+p_y^2}{m}+2\ell^2\zeta-mgy=0 \right\}\,,\]
and the Hamiltonian vector field describing the dynamics is
\begin{align*}
X_x&=\frac{p_x}{m}\,, &X_y&=\frac{p_y}{m}\,,  &X_\zeta&=\frac{3g}{2\ell^2}p_y\,,\\
X_{p_x}&=2\zeta x\,,       &X_{p_y}&=2\zeta y-mg\,,     &X_{p_\zeta}&=0\,.
\end{align*}

\medskip

\noindent \textbf{Remarks}
\begin{enumerate}
\item It is instructive to compare the preceding computation with the one obtained from the Dirac algorithm in its original form (using Poisson brackets and so on). It can be seen that there is a perfect match between the intermediate results of both procedures i.e.\ how the primary and secondary constraints appear and how the value of $u$ is fixed. This notwithstanding, for field theories in not very familiar settings (for instance, when boundaries are present) the approach that we have followed above is far more transparent and less prone to errors than Dirac's method. This is, in fact, one of the main points of the paper.
\item The equations for the integral curves of the Hamiltonian vector field $X$ are \emph{strictly equivalent} to the Euler-Lagrange equations \eqref{eulerpendulum}. The easiest way to show this is to find a parametrization of the constraints defining the submanifold $\mathcal{C}$ (for instance $x=\ell\sin\theta$, $y=-\ell\cos\theta$, $p_x=\pi_\theta\cos\theta$ and $p_x=\pi_\theta\sin\theta$ with $\theta\in[0,2\pi)$, $\pi_\theta\in \mathbb{R}$ and solving for $\zeta$ in $\phi_4=0$) and writing the equations of the integral curves in terms of these variables. By doing this one gets, precisely, equations (\ref{eq.1}) and (\ref{eq.2}).
\item The Dirac multiplier $u$ has been completely fixed. Its time dependence originates in the one of $p_y$.  The role of the Dirac multiplier is to enforce the primary constraint $p_\zeta=0$ as a subsidiary condition that must be satisfied at all times, this is the reason why it is included in the action as discussed above. Its role can also be seen from a different (more geometric) perspective. The Hamiltonian $H$ is only fixed on the primary constraint submanifold $\mathcal{M}_0$ and, hence, it admits many different extensions. With the help of $u$ it is possible to have a sufficiently large family of Hamiltonians providing Hamiltonian vector fields that can be adjusted (by appropriately choosing $u$) in such a way that we get consistent dynamics.
\item It is important to highlight the difference between the roles of $u(t)$ and the configuration variable $\zeta$. Despite the fact that, superficially, $\zeta$ appears to be the same type of object as $u$, its role is different. First of all, $\zeta$ is part of the configuration space for the system. Second, it has an interesting dynamical interpretation as it is possible to find the tension of the pendulum bar from the knowledge of $\zeta$. It is introduced in the Lagrangian as a way to obtain the condition $x^2+y^2=\ell^2$ \emph{dynamically} as an equation of motion. It should be mentioned here that the lapse and the shift in the Hamiltonian description of general relativity appear in a very similar way (although they play quite a different physical role).
\item A final comment that will be very important for the field theories that we discuss in the rest of the paper. In the preceding finite dimensional model we have worked in particular charts that sufficed to cover the full configuration space $\mathcal{Q}$, and its tangent and cotangent bundles $T\mathcal{Q}$ and $T^*\mathcal{Q}$. In particular, we have written the canonical symplectic form as \eqref{symplecticpendulum} and vector fields as \eqref{vector_field}. An alternative way of looking at $\Omega$ is to consider its action on a pair of vector fields on phase space $X,Y\in\mathfrak{X}(T^*\mathcal{Q})$
\begin{equation*}
\Omega(X,Y)=Y_{p_x}X_{x}  -X_{p_x}Y_{x}+Y_{p_y}X_{y}-X_{p_y}Y_{y}+Y_{p_\zeta}X_{\zeta}-X_{p_\zeta}Y_{\zeta}\,.
\end{equation*}
As we will show in the following, this way of writing the symplectic form will be instrumental to avoid many of the problems that crop up when studying field theories in the presence of boundaries.\label{remark Omega vector fields}\newcounter{aux}\setcounter{aux}{\value{enumi}}
\end{enumerate}

\section{The scalar field with Dirichlet boundary conditions}\label{Sect_Dirichlet}

We study now the first field model: a massless scalar field $\varphi$ defined in a bounded region $\Sigma$ and subject to Dirichlet boundary conditions\footnote{It is straightforward to extend the results of this section to arbitrary, time-independent, Dirichlet boundary conditions.} $\varphi|\partial\Sigma=0$. For the sake of simplicity we will concentrate on a $(1+1)$-dimensional example. Our results and methods complement those obtained in \cite{BPV1} with the help of the GNH algorithm and should be compared with those presented in \cite{Jabbari}. Notice that in \cite{BPV1} the Dirichlet boundary conditions were introduced, from the start, in the definition of the configuration space whereas in the following discussion they will appear as natural boundary conditions (i.e.\ as a consequence of the field equations) for a modified Lagrangian for the scalar field.

Let us take a space $\mathcal{F}$ of sufficiently smooth real functions on the interval $\Sigma=[0,1]$. In principle, it suffices to consider elements $\varphi$ of this space to represent the scalar field configurations. In order to incorporate the Dirichlet boundary conditions we will introduce a second, auxiliary, scalar field $\chi$. Our configuration space will be of the form $\mathcal{Q}=\mathcal{F}\times\mathcal{F}$ and  $T\mathcal{Q}\cong(\mathcal{F}\times\mathcal{F})\times(\mathcal{F}\times\mathcal{F})$. Let us consider the Lagrangian
\begin{equation}
\label{Lagrangian_Dirichlet}
L:T\mathcal{Q}\rightarrow \mathbb{R}:\mathrm{v} \mapsto L(\mathrm{v})=\frac{1}{2} \int_\Sigma  \left(v_\varphi^2 - \varphi'^2  +2\left( \chi \varphi \right)' \right)\,,\\
\end{equation}
where we denote $\mathrm{v}:=(\varphi, \chi ; v_\varphi, v_{\chi})$ and use the prime to represent the spatial derivative. Here and in the following we omit the measure $\mathrm{d}x$ in the integrals. The Lagrangian \eqref{Lagrangian_Dirichlet} can, obviously, be written as\footnote{The idea of introducing a surface term via a total derivative to avoid complications in the computation of Poisson brackets bears an interesting resemblance with the proposal presented in \cite{Perez} to understand the holonomy-flux algebra.}
\begin{equation}
\label{Lagrangian_Dirichlet_boundary}
L(\mathrm{v})=\chi(1)\varphi(1)-\chi(0)\varphi(0)+\frac{1}{2}\int_\Sigma \left(v_\varphi^2-\varphi'^2\right)\,,
\end{equation}
and, hence, the role of the field $\chi$ is clear: its boundary values $\chi(0)$ and $\chi(1)$ enforce\footnote{It is interesting to ponder at this point to what extent the field values at the boundary points of $\Sigma$ can be thought of as ``degrees of freedom'' as they are fixed by continuity by the values of the fields in the interior of $\Sigma$.} the conditions $\varphi(0)=\varphi(1)=0$.

The Euler-Lagrange equations for \eqref{Lagrangian_Dirichlet} or \eqref{Lagrangian_Dirichlet_boundary} are
\begin{align*}
\ddot{\varphi}-\varphi''=0\,,&&&  \textrm{on}\, (t_1,t_2)\times[0,1]\,,\\
\varphi(0)=\varphi(1)=0\,,&&&  \forall t\in(t_1,t_2)\,,\\
\chi(1)=\varphi'(1)\,,\chi(0)=\varphi'(0)\,,&&&  \forall t\in(t_1,t_2)\,.
\end{align*}
As we can see, we get the wave equation for the scalar field $\varphi$ subject to homogeneous Dirichlet boundary conditions. For all $t\in(t_1,t_2)$ the values of $\chi(0)$ and $\chi(1)$ are given by the derivatives of $\varphi$ at the boundary points of $\Sigma$, but $\chi$ is, otherwise, left arbitrary. Although, strictly speaking, we have enlarged our model by the addition of the extra field $\chi$, there is a straightforward one to one correspondence between the solutions to the model that we are considering here and the ones for the usual free scalar on the interval $\Sigma$ subject to homogeneous Dirichlet boundary conditions.

If we take $\mathrm{v}, \mathrm{w}\in T_{(\varphi,\chi)} \mathcal{Q}$, $\mathrm{v}:=(\varphi,\chi;v_\varphi,v_\chi)$, $\mathrm{w}:=(\varphi,\chi;w_\varphi,w_\chi)$ the fiber derivative is
\begin{eqnarray}\label{momento}
\langle F\!L(\bf{\mathrm{v}})|\bf{\mathrm{w}} \rangle&=&\int_{\Sigma} v_\varphi w_\varphi,\quad\longrightarrow\quad
\begin{array}{l}
{\bf p}_{\varphi}(\cdot):=\displaystyle \int_{\Sigma}  v_\varphi\,\cdot \,,\\
\quad\\
{\bf p}_{\chi} (\cdot):=0 \,.
\end{array}
\end{eqnarray}
We use bold letters to denote elements in the dual space $\mathcal{F}^*$. We will assume\footnote{This is another instance where functional analytic issues are relevant.} that all the momenta can be written as integrals over the interval $[0,1]$
\[
{\bf p}_{\varphi}(\cdot):=\displaystyle \int_{\Sigma}  p_\varphi\,\cdot
\]
with $p_\varphi\in\mathcal{F}$ hence, in the following we will represent the momenta ${\bf p}_{\varphi}(\cdot)$ as $p_\varphi$ if convenient.

As we can see, we get as primary constraint the condition that the momentum canonically conjugate to the auxiliary field $\chi$ vanishes. The energy is
\begin{eqnarray*}
E=\langle F\!L({\rm{v}})|{\rm{v}}\rangle -L\left(\rm{v} \right)=  \frac{1}{2} \int_{\Sigma}  \left(v_{\varphi}^2 +\varphi'^2 -2\left( \chi \varphi \right)' \right)\,,
\end{eqnarray*}
and an appropriate extension of the Hamiltonian to the full phase space $T^*\mathcal{Q}$ is
\begin{equation}\label{HamiltonianDir}
H= \frac{1}{2} \int_{\Sigma}  \left( p_{\varphi}^2+\varphi'^2 -2\left( \chi \varphi \right)' \right)\,,
\end{equation}
A generic vector field on $T^*\mathcal{Q}$ consists of vectors $Y \in T_{\left( \varphi, \chi ; p_{\varphi},p _{\chi} \right)}T^*\mathcal{Q}$ of the form
\begin{equation*}
Y= \left( \left( \varphi, \chi ; p_{\varphi},p _{\chi} \right), ( Y_{\varphi}, Y_{\chi}, \boldsymbol{Y}_{{\bf p}_{\varphi}}(\cdot), \boldsymbol{Y}_{{\bf p}_{\chi}}(\cdot))\right)\,,
\end{equation*}
where  $Y_{\varphi}, Y_{\chi} \in \mathcal{F}$. The components $\boldsymbol{Y}_{{\bf p}_{\varphi}}(\cdot), \boldsymbol{Y}_{{\bf p}_{\chi}}(\cdot)$ can be represented by real functions\footnote{They can be thought of as elements of the dual $\mathcal{F}^*$.} $Y_{{\bf p}_{\varphi}},Y_{{\bf p}_{\chi}}$ such that over $f,g\in\mathcal{F}$ we have
\[
\boldsymbol{Y}_{{\bf p}_{\varphi}}(f):=\int_\Sigma Y_{{\bf p}_{\varphi}} f\,,\quad\boldsymbol{Y}_{{\bf p}_{\chi}}(g):=\int_\Sigma Y_{{\bf p}_{\chi}}g\,.
\]
For this reason, in the following we will represent vector fields as
\[
Y= \left( \left( \varphi, \chi ; p_{\varphi},p _{\chi} \right), ( Y_{\varphi}, Y_{\chi}, Y_{{\bf p}_{\varphi}}, Y_{{\bf p}_{\chi}})\right)\,.
\]
When acting on a vector field $Y$ the exterior differential of the Hamiltonian is
\begin{equation}
\ed H(Y)=\left[ \left(\chi-\varphi'\right)Y_{\varphi}+\varphi Y_{\chi}\right](0)-\left[\left(\chi-\varphi'\right)Y_{\varphi}+\varphi Y_{\chi}\right] (1)+\!\!\int_\Sigma\!\left( Y_{{\bf p}_{\varphi}} p_{\varphi} - \varphi''  Y_{\varphi}\right)\,.\label{diffHamiltonianDir}
\end{equation}
The  canonical symplectic form in $T^*\mathcal{Q}$,  acting on a pair of vector fields $X,Y$ defined on the phase space (see remark \ref{remark Omega vector fields}), is
\begin{align}\label{OmegaXY}
\begin{split}
\Omega(X,Y)
&= \boldsymbol{Y}_{{\bf p}_{\varphi}}\!\left(X_{\varphi}\right)- \boldsymbol{X}_{{\bf p}_{\varphi}}\!\left( Y_{\varphi} \right) +\boldsymbol{Y}_{{\bf p}_{\chi}}\!\left(X_{\chi}\right)- \boldsymbol{X}_{{\bf p}_{\chi}}\!\left( Y_{\chi} \right)\\
&=\int_\Sigma \left( Y_{{\bf p}_{\varphi}} X_{\varphi}  - X_{{\bf p}_{\varphi}} Y_{\varphi} + Y_{{\bf p}_{\chi}} X_{\chi}  - X_{{\bf p}_{\chi}} Y_{\chi} \right)\,.
\end{split}
\end{align}
Let us look now for vector fields $X$ satisfying the equation (see eq.\ \eqref{ecs_X_u} where now $\lambda$ is the Dirac multiplier introduced to enforce the primary constraints at the level of the action)
\begin{equation}\label{EcOmegaEscalar}
\Omega(X,Y)= \ed H(Y)+\langle {\bf \lambda}| \ed p_{\chi}\rangle(Y)\,,
\end{equation}
for \emph{every} vector field $Y$. The last term means
\[
\langle {\bf \lambda}| \ed p_{\chi}\rangle(Y):=\int_\Sigma \lambda Y_{{\bf p}_\chi}\,.
\]
By considering first those fields $Y$ vanishing at the boundary of the interval $[0,1]$ we get the following expressions for the Hamiltonian vector field $X$ in the whole interval $[0,1]$
\begin{equation}\label{HVSFD}
\begin{alignedat}{3}
X_{\varphi}&=  p_{\varphi}\,,    \qquad\qquad\qquad\qquad &  X_{\chi}&=\lambda\,,\\
X_{{\bf p}_{\varphi}}&= \varphi''\,, &  X_{{\bf p}_{\chi}} &=0\,.
\end{alignedat}
\end{equation}
Once we know $X$, we can allow $Y$ to be arbitrary on the boundary. This gives us, then, the following additional secondary constraints
\begin{subequations}
\begin{align}
\varphi(0)&=0 &\varphi(1)&=0\,,\label{bc1}\\
\chi(0)-\varphi'(0) &=0 &\chi(1)-\varphi'(1)&=0\,,\label{bc2}
\end{align}
\end{subequations}
which include both the Dirichlet boundary conditions and the values of $\chi$ at the boundary. Notice that we find the result given by the Euler-Lagrange equations.

Now, we must check the tangency of the Hamiltonian field, \eqref{HVSFD}, to the submanifold in $T^*\mathcal{Q}$ defined by the constraints $p_{\chi}=0$ and the boundary conditions eq.\ \eqref{bc1},\eqref{bc2}. A straightforward computation gives
\begin{subequations}
\begin{align}
&0=\bm{\imath}_{X} \ed  p_{\chi}=X_{{\bf p}_{\chi}} \,, \label{e1}\\
&0=\bm{\imath}_{X} \ed \left(\varphi(j)\right)=X_\varphi(j)= p_{\varphi} (j)\,,\quad j\in\{0,1\}\,, \label{e2}\\
&0=\bm{\imath}_{X} \ed \left( \chi(j)-\varphi'(j)\right)=X_\chi(j)-X'_\varphi(j)= \lambda(j)-p_{\varphi}'(j)\,, \quad j\in\{0,1\}\,. \label{e4}
\end{align}
\end{subequations}
As we can see, the tangency condition \eqref{e1} gives nothing new, the next pair of conditions \eqref{e2} are new secondary constraints on the values of the momenta $p_\varphi$ at the boundary of $\Sigma$. Finally, \eqref{e4} fixes the Dirac multiplier at the boundary $\lambda(0)=p_{\varphi}'(0)$, $\lambda(1)=p_{\varphi}'(1)$.

We must demand now that the vector field $X$ be tangent to the new submanifold defined by the secondary constraints just obtained. These new tangency conditions give
\begin{align}
0&=\bm{\imath}_{X} \ed\left(p_\varphi(j)\right)=X_{{\bf p}_\varphi}(j)=D^2 \varphi(j)\,,\quad j\in\{0,1\}\,,
\end{align}
where $D^n$ denotes the $n$-th order spatial derivative. As we see, there are more secondary constraints and additional tangency requirements. Iterating this process, we find an infinite number of boundary constraints of the form ($n\in\mathbb{N}$)
\begin{equation}\label{boundary_constraints}
\begin{alignedat}{3}
D^{2n} \varphi (0)&=0\,,    \qquad\qquad\qquad &  D^{2n} p_{\varphi} (0)&=0\,,\\
D^{2n} \varphi (1)&=0\,, &  D^{2n} p_{\varphi} (1)&=0\,.
\end{alignedat}
\end{equation}

\medskip

\noindent \textbf{Remarks}
\begin{enumerate}\setcounter{enumi}{\value{aux}}
\item The Dirac multiplier $\lambda$ (a real continuous function on $[0,1]$) is fixed only at the boundary of $\Sigma$. Its value in $(0,1)$ is otherwise arbitrary. This arbitrariness implies that, under the dynamical evolution of the system, $\chi$ is arbitrary in $(0,1)$, but its evolution is tied to that of $\varphi'$ at the boundary points. Strictly speaking, this is a gauge invariance of the system similar to the one that we will encounter when discussing the Hamiltonian formulation for the Pontryagin action. Notice, however, that it only involves the auxiliary field $\chi$.
\item In practical terms, $\bm{\imath}_X \ed$ should be thought of as a directional derivative in phase space. Hence, requiring that the Hamiltonian vector field be tangent to the phase space submanifold defined by a condition of the type $\Phi_n=0$ simply amounts to demanding $\bm{\imath}_X \ed \Phi_n=0$. This is so irrespective of the geometric nature of $\Phi_n$ (a section of some tensor bundle over $\Sigma$), i.e.\ of the tensorial character of the constraints in their standard interpretation as fields on $\Sigma$.
\item The constraints \eqref{boundary_constraints} are a well-known feature of the solutions to the wave equation (see, for instance, \cite[p.327]{Brezis}). Their actual number depends on the regularity demanded of the solutions to the field equations. As we are formally allowing for as much smoothness as we wish, in the present case we get an infinite tower of them. Their presence is to be expected as the time derivatives of any order of $\varphi(0)$ and $\varphi(1)$ must vanish. As these time derivatives can be written in terms of the spatial ones by using the wave equation in $[0,1]$, we get conditions on the even order derivatives of both $\varphi$ and $p_\varphi$ at $x\in\{0,1\}$.  A direct way to see why these constraints are related to the smoothness of the solutions is to use the d'Alembert formula (valid for general solutions to the wave equation in $1+1$ dimensions). Indeed, consider some smooth initial data with $\varphi(0,x)=:f(x)$ and $\varphi'(0,x)=:g(x)$. In order to obtain solutions in $[0,1]$ subject to Dirichlet boundary conditions, $f$ and $g$ must be extended as odd functions to the whole $\mathbb{R}$. These extensions will be smooth iff the even order derivatives of $f$ and $g$ vanish at $x=0$ and $x=1$.
\item Other types of boundary conditions (for instance, Robin) can be dealt with in a similar way by introducing them in the action with the help of appropriate surface terms (as was done in \cite{BPV1} within the GNH framework).
\item The advantage of working with \eqref{Lagrangian_Dirichlet} instead of \eqref{Lagrangian_Dirichlet_boundary} is that it makes it unnecessary to use continuous and discrete  configuration variables at the same time ($\varphi$ and $\chi(0)$, $\chi(1)$ respectively). This is specially important when writing the canonical symplectic form and effectively eliminates some of the ambiguities that may crop up when formally following the Dirac method in the presence of spatial boundaries \cite{Jabbari}. These problems can, of course, be avoided by carefully taking into account functional analytic details but we feel that the rule of thumb that we are suggesting in this example---replace, when possible, surface terms in the action by total derivatives---can be instrumental to avoid making mistakes.
\item By directly looking for the solutions to \eqref{EcOmegaEscalar}, we have been able to avoid the explicit computation of Poisson brackets. In our opinion, this is the main advantage of proceeding in the way that we have explained here. If one tries to blindly use the Dirac approach and the standard writing the canonical Poisson brackets as $\{\varphi(x),p_\varphi(y)\}=\delta(x,y)$, one quickly faces the conundrum of making sense of expressions such us $\{\varphi(0),p_\varphi(0)\}$. Although it may be possible \emph{a posteriori} (i.e.\ once we know what we have to get) to come up with heuristic ways to deal with such objects, the advantages of following a systematic procedure, such as the one used here, are clear.\setcounter{aux}{\value{enumi}}
\end{enumerate}

\section{Abelian Chern-Simons without boundaries}\label{sec_CS}

We discuss now the treatment of the abelian\footnote{The extension of our analysis to the non-abelian case is straightforward. For our purposes here it suffices to consider the abelian Chern-Simons model.} Chern-Simons model in a 3-dimensional manifold\footnote{Strictly speaking, we should consider a manifold of the form $\mathcal{B}=[t_1,t_2]\times\Sigma_2$ and consider field variations vanishing on $\{t_1\}\times\Sigma_2$ and $\{t_2\}\times\Sigma_2$.}   $\mathcal{B}=\mathbb{R}\times\Sigma_2$ (where $\Sigma_2$ denotes a 2-dimensional manifold without boundary). Let us consider the action
\begin{equation}
S_{\mathrm{CS}}(a)=\int_{\mathcal{B}} a \wedge F(a)\,,  \label{CSNN}
\end{equation}
where $F(a)=\mathrm{d}a$ is the curvature of the three-dimensional 1-form $a$. The Euler-Lagrange equations are simply
\begin{equation}
F(a)=\mathrm{d}a=0\,,
\end{equation}
and their solutions are the closed 1-forms on $\mathcal{B}$.

In the present case (as it happens also in general relativity), we are not starting from a Lagrangian but directly from the action. In order to find a Lagrangian for the model it is necessary to perform a 2+1 decomposition. We do this by expanding $a=a_{\perp }\mathrm{d}t +a^\top$, so that $F=\mathrm{d}a=\mathrm{d}t\wedge F_{\perp}+\mathrm{d}^\top a^\top$, where\footnote{$\pounds_{\bf{t}}$ denotes the Lie derivative along the field $\bf{t}$ and $\mathrm{d}^\top$ is the exterior differential on $\Sigma_2$.} $F_{\perp}:=\pounds_{{\bf{t}}}a^{\top}-\mathrm{d}^\top a_{\perp}$. Then, the action \eqref{CSNN} can be written as
\begin{equation}
S_{\mathrm{CS}}(a)=\int_{\mathbb{R}}\mathrm{d}t\int_{\Sigma_2}\left(\pounds_{{\bf{t}}} a^{\top}\wedge a^{\top}+ a^{\top}\wedge\mathrm{d}^\top a_{\perp}+a_{\perp}\mathrm{d}^\top\!a^\top\right). \label{DCS}
\end{equation}
In order to simplify the notation, we write in the following $q_{\perp}:=a_{\perp}$, $q:= a^\top$, and $\mathrm{d}:=\mathrm{d}^\top$. From \eqref{DCS} we can read off the Lagrangian
\[\begin{array}{cccc}
L: &  T\mathcal{Q}:= T\left( \Omega^0(\Sigma_2)\times \Omega^1(\Sigma_2) \right)  & \longrightarrow & \R \\
 &          \quad\quad\textrm{v}:=(q_{\perp},q;v_{\perp},v)  & \longmapsto & L(\textrm{v} )
\end{array} \]
which is given by
\begin{equation}\label{Lagrangian_CS}
L(\textrm{v})=\int_{\Sigma_2}\left(v\wedge q+ q\wedge\mathrm{d}q_{\perp}+q_{\perp}\mathrm{d}q\right).
\end{equation}

\medskip

\noindent The fiber derivative $F\!L:T \mathcal{Q}\to T^*\mathcal{Q} $ is now
\[
\begin{array}{l}
\displaystyle
\left\langle F\!L\left(\textrm{v}\right)| \textrm{w} \right\rangle=\int_{\Sigma_2} w \wedge q
\end{array}
\quad \longrightarrow\quad
\begin{array}{l}
{\bf p}_{\perp}\left(\cdot\right):=0\,,\\[1.7ex]
{\bf p}\left(\cdot\right)\,:=\displaystyle \int_{\Sigma_2} \cdot \wedge q\,,
\end{array}
\]
where $\left( {\bf p}_{\perp}, {\bf p}\right)$  are elements of the dual of the configuration space (linear functionals). This means that acting on functions $f\in\Omega^0(\Sigma_2)$ and 1-forms $\alpha\in\Omega^1(\Sigma_2)$ they give
\[
{\bf p}_{\perp}(f)=0\,, \qquad\qquad {\bf p}(\alpha)=\int_{\Sigma_2}\alpha\wedge q\,.
\]
The definition of the momenta tells us that we have the following primary constraints
\begin{align}\label{const_CS}
{\bf C}_{\perp}\left(\cdot\right):= {\bf p}_{\perp}\left(\cdot\right)=0\,,\qquad\qquad {\bf C}(\cdot):= {\bf p}(\cdot)-\int_{\Sigma_2} \cdot \wedge q=0\,.
\end{align}
On the primary constraint submanifold the Hamiltonian is
\begin{equation}\label{Hamiltonian_CS}
H  
=\int_{\Sigma_2}\big(\mathrm{d}q_\perp\wedge q-q_\perp\mathrm{d}q\big)\,.
\end{equation}
As $H$ depends only on configuration variables, it can be immediately extended to a function defined on the whole phase space.

Vector fields $X\in\mathfrak{X}(T^*\mathcal{Q})$ have the form
\begin{equation}\label{Vector_field_CS}
X=\big(\left(q_{\perp},q,{\bf p}_{\perp},{\bf p}\right), \left(X_{q_{\perp}},X_{q},{\bf X}_{{\bf p}_{\perp}},{\bf X}_{{\bf p}}\right)\big)\,.
\end{equation}
Notice that the geometric interpretation of the different components is important: $X_{q_\perp}\in \Omega^0(\Sigma_2)$, $X_q\in \Omega^1(\Sigma_2)$, whereas ${\bf X}_{{\bf p}_{\perp}}$ and ${\bf X}_{{\bf p}}$ are elements of their duals.

\medskip

The canonical symplectic form on the cotangent bundle $T^*\mathcal{Q}$  acting on a pair of vector fields $(X,Y)$ is in the present case:
\begin{equation}\label{OmegaXY_CS}
\Omega(X,Y)={\bf Y}_{{\bf p}_{\perp}}\!\left(X_{q_{\perp}}\right)-{\bf X}_{{\bf p}_{\perp}}\!\left( Y_{q_{\perp}}\right)+{\bf Y}_{{\bf p}}\!\left( X_{q}\right)-{\bf X}_{{\bf p}}\!\left( Y_{q}\right)\,.
\end{equation}

\noindent Next, we have to solve the equation

\begin{equation}\label{HFCS}
\Omega(X,Y)=\ed H \left(Y\right) +\langle\lambda_{\perp}|\ed{\bf C}_{\perp}\rangle \left(Y\right)+\langle\lambda|\ed{\bf C}\rangle\left(Y\right)
\end{equation}

\bigskip

\noindent for any arbitrary $Y$ vector field and Dirac multipliers $\lambda_{\perp} \in \Omega^0\left(\Sigma_2\right)$, $\lambda\in \Omega^1\left(\Sigma_2\right)$. The right-hand side of \eqref{HFCS} is
\begin{equation}
{\bf Y}_{{\bf p}_{\perp}}\!\left( \lambda_{\perp}\right)-\int_{\Sigma_2} \left[ Y_{q_{\perp}} \mathrm{d} q- \mathrm{d} Y_{q_{\perp}}\wedge q \right]
+{\bf Y}_{{\bf p}}\!\left( \lambda \right)-\int_{\Sigma_2} \left[ q_{\perp} \mathrm{d} Y_{q}+ Y_{q}\wedge\mathrm{d}q_{\perp} -Y_{q}\wedge \lambda \right]\,,
\end{equation}
therefore,  we obtain the Hamiltonian vector field $X$
\begin{align}\label{NHVFCS}
 \begin{alignedat}{3}X_{q_{\perp}}&=\lambda_{\perp}\,,\qquad\qquad &{\bf X}_{{\bf p}_{\perp}}\!\left(\cdot\right)&=2\int_{\Sigma_2}\cdot \,{{\rm{d}}}q\,,\\
 X_{q}&= \lambda\,,                &{\bf X}_{{\bf p}}\!\left(\cdot\right)        &=\int_{\Sigma_2} \cdot \wedge \left(2{\rm{d}}q_{\perp}-\lambda\right)\,,
 \end{alignedat}
\end{align}
where we have made use of the fact that the boundary of $\Sigma_2$ is empty. The tangency conditions on the primary constraints give
\begin{align*}
&0=\bm{\imath}_X \ed \left( {\bf C}_{\perp} (\cdot)\right)={\bf X}_{{\bf p}_{\perp}}(\cdot)=2\!\!\int_{\Sigma_2}\!\!\!\cdot \,\mathrm{d}q\,,\\
&0=\bm{\imath}_X  \ed \left( {\bf C}(\cdot)\right)={\bf X}_{{\bf p}}(\cdot)-\!\int_{\Sigma_2}\cdot \wedge X_q=2\!\!\int_{\Sigma_2}\!\!\!\cdot\wedge \left(\mathrm{d}q_{\perp}-\lambda\right)\,,
\end{align*}
hence, we have a secondary constraint $\mathrm{d} q=0$ and the Dirac multiplier is fixed to be $\lambda= \mathrm{d} q_{\perp}$. The tangency condition on the new secondary constraint gives
\begin{equation*}
0=\bm{\imath}_X  {\ed} \left( {{\rm{d}}} q \right)= {{\rm{d}}} X_q= {{\rm{d}}} \lambda =\mathrm{d}^2q_{\perp}=0\,,
\end{equation*}
As we can see, there are no further constraints, and the algorithm stops. The final constraint submanifold is
\begin{equation}\label{final_Constr_CS}
\mathcal{C}_{CS}:=\big\{  \left( q_{\perp}, q, {\bf p}_{\perp}, {\bf p} \right) \in T^*\mathcal{Q} : {\bf p}_{\perp}(\cdot)=0,\, {\bf p}(\cdot)-\int_{\Sigma_2} \cdot \wedge q = 0, \,    {\rm{d}}q=0\big\}\,
\end{equation}
and the Hamiltonian vector field
\begin{align}\label{HVFCSG}
\begin{alignedat}{3}
X_{q_{\perp}}&=\lambda_{\perp}, \qquad\qquad\qquad\qquad& {\bf X}_{{\bf p}_{\perp}}\left(\cdot\right)&=0\,,\\
X_{q}&={\rm{d}} q_{\perp},& {\bf X}_{{\bf p}}  \left(\cdot\right)&= \int_{\Sigma_2} \cdot \wedge {\rm{d}} q_{\perp}\,.
\end{alignedat}
\end{align}

\noindent \textbf{Remarks}
\begin{enumerate}\setcounter{enumi}{\value{aux}}
\item The interpretation of the Hamiltonian vector field $X$ and the constraints is straightforward. The field equation $F(a)=0$ is now a consequence of the (secondary) constraint ${\rm{d}}q=0$ and the ``evolution'' equation $\dot{q}-{\rm{d}}q_\perp=0$. The presence of the arbitrary function of time $\lambda_\perp(t)$ in the Hamiltonian vector field can be immediately interpreted as the 2+1 abelian gauge symmetry $a\mapsto a+{\rm{d}}\mu$.

\item Notice that different extensions of the Hamiltonian produce different intermediate results when following the procedure that we are proposing. Of course, they lead to dynamically equivalent forms for the final constraint submanifold, Hamiltonian vector fields, dynamics, and gauge symmetries.

\item The dynamics of the configuration variables can be obtained directly from the $X_{q_{\perp}}$, $X_{q}$ components of the Hamiltonian vector field $X$. Notice, in particular, that the momenta play no role in this (something to be expected because the dynamics is given by the \emph{first order} differential equation $\mathrm{d}a=0$).

\item In the preceding analysis the Dirac multiplier $\lambda$ is fixed whereas $\lambda_\perp$ is left arbitrary. This signals the presence of second and first class constraints.\footnote{The geometric interpretation of the final constraint submanifold---its first or second class character---can be found in the standard way so we refrain to do it here.}\setcounter{aux}{\value{enumi}}

\end{enumerate}

\section{Pontryagin with boundaries}\label{sec_Pontryagin}

We will study now the abelian Pontryagin model defined on a 4-dimensional manifold $\mathcal{M}=\mathbb{R}\times\Sigma_3$ where $\Sigma_3$ is allowed to have boundary. Let us then consider the action
\begin{equation}\label{action_Pontryagin}
  S_{Pg}(A)=\int_{\mathcal{M}} F(A)\wedge F(A)\,,
\end{equation}
where $F(A):=\mathrm{d}A$ is the curvature of a four dimensional 1-form $A$. Performing an integration by parts, this action can be written as
\begin{equation}\label{action_Pontryagin_2}
  S_{Pg}(A)=\int_{\mathbb{R}\times\partial\Sigma_3} a\wedge da\,,
\end{equation}
where $a$ denotes the pullback of the connection $A$ to $\partial\mathcal{M}=\mathbb{R}\times\partial\Sigma_3$, the boundary of $\mathcal{M}$. The dynamics of the model defined by \eqref{action_Pontryagin} is then easy to interpret: the solutions to the Euler-Lagrange equations are the $U(1)$ connections on $\mathcal{M}$ with flat pullbacks to the boundary $\partial\mathcal{M}$ (i.e.\ ${\mathrm{d}}a=\mathrm{d}(\imath^*_\partial A)=0$).

The Lagrangian in the present case can be obtained by performing a 3+1 decomposition analogous to the one used in section \ref{sec_CS} (writing  $A=A_{\perp}{\rm{d}}t+A^\top$), the  result is
\[\begin{array}{cccc}
L: &  T\mathcal{Q}:= T\left( \Omega^0(\Sigma_3)\times \Omega^1(\Sigma_3) \right)  & \longrightarrow & \R \\
 &          \mathrm{V}=(Q_{\perp},Q;V_{\perp},V)  & \longmapsto & \displaystyle 2\int_{\Sigma_3}  \left(V-{\rm{d}} Q_{\perp} \right) \wedge {\rm{d}} Q
\end{array} \]
where now $Q_{\perp}:=A_\perp$ and $Q:=A^\top$. The fiber derivative is
\[
\begin{array}{l}
\displaystyle \left\langle F\!L \left(\mathrm{V}\right)|\mathrm{W}\right\rangle= 2\int_{\Sigma_3} W\wedge \mathrm{d}Q
\end{array}
\quad \longrightarrow
\begin{array}{l}
\displaystyle {\bf P}_{\perp}\left(\cdot\right) := 0\,, \\[1.7ex]
\displaystyle {\bf P}\left(\cdot\right):= 2\displaystyle \int_{\Sigma_3} \cdot \wedge {\mathrm{d}} Q\,,
\end{array}
\]
where, $\mathrm{V}, \mathrm{W} \in T_{(Q_{\perp},Q)}\mathcal{Q}$, $\mathrm{V}:=(Q_{\perp},Q;V_{\perp},V), \mathrm{W}:=(Q_{\perp},Q;W_{\perp},W)$. From the definition of the momenta we get the following  primary
constraints
\[\begin{array}{rlcrl}\label{conspPG}
{\bf C}_{\perp}\left(\cdot\right):={\bf P}_{\perp}\left(\cdot\right) =0\,,\quad{\bf C}\left(\cdot\right):={\bf P}\left(\cdot\right)- 2\displaystyle\int_{\Sigma_3} \cdot \wedge {\mathrm{d}} Q  = 0\,.
\end{array} \]
The Hamiltonian is
\begin{equation}\label{Hamiltonian_PG}
H
=2\int_{\Sigma_3}{\mathrm{d}}Q_{\perp}\wedge {\mathrm{d}}Q\,,
\end{equation}
and the vector fields $X\in\mathfrak{X}(T^*\mathcal{Q})$ have now the form
\begin{equation}\label{Vector_field_PG}
X=\big( \left( Q_{\perp}, Q, {\bf P}_{\perp}, {\bf P} \right), \left( X_{Q_{\perp}}, X_{Q}, {\bf X}_{{\bf P}_{\perp}}, {\bf X}_{{\bf P}} \right) \big)\,.
\end{equation}
The canonical symplectic form on the cotangent bundle $T^*\mathcal{Q}$  acting on pairs of vector fields is
\begin{equation}\label{OmegaXY_PG}
\Omega(X,Y)={\bf Y}_{{\bf P}_{\perp}}\!\left( X_{Q_{\perp}}\right)-{\bf X}_{{\bf P}_{\perp}}\!\left( Y_{Q_{\perp}}\right)+{\bf Y}_{{\bf P}}\!\left( X_{Q}\right) - {\bf
X}_{{\bf P}}\!\left( Y_{Q}\right)\,.
\end{equation}
In order to proceed we have to solve the equation
\begin{equation}\label{HFP}
\Omega(X,Y)=\ed H (Y) +\langle \Lambda_{\perp}| \ed{\bf C}_{\perp}\rangle\left(Y\right)+ \langle\Lambda|\ed{\bf C}\rangle\left(Y\right)\,.
\end{equation}
The right hand side of \eqref{HFP} can be written as
\begin{equation}\label{rhs_FP}
{\bf Y}_{{\bf P}_{\perp}}\!\left( \Lambda_{\perp}\right)+2\int_{\Sigma_3} {\mathrm{d}} Y_{Q_{\perp}} \wedge{\mathrm{d}} Q
+{\bf Y}_{{\bf P}}\!\left(\Lambda \right) - 2\int_{\Sigma_3}  \left(\Lambda-{\mathrm{d}} {Q}_{\perp}\right) \wedge {\mathrm{d}} Y_Q\,.
\end{equation}
Therefore,  we obtain the Hamiltonian vector field with components
\begin{align}\label{NHVFPG1}
\begin{alignedat}{3}
 X_{Q_{\perp}}  &=  \Lambda_{\perp}\,, \qquad\qquad\qquad&  { \bf X}_{{\bf P}_{\perp}}  \left(\cdot\right) &= - 2\int_{\partial\Sigma_3} \imath_{\partial}^*(\cdot{}){\mathrm{d}}^{\partial} Q^{\partial} \,,\\
 X_{Q}          &=\Lambda\,,           & {\bf X}_{{\bf P}}  \left(\cdot\right)           &= 2\int_{\Sigma_3} \cdot \wedge {\mathrm{d}} \Lambda +2\int_{\partial\Sigma_3} \imath_{\partial}^*(\cdot{})\wedge \left(\Lambda-{\mathrm{d}} {Q}_{\perp}\right)^{\partial} \,,
 \end{alignedat}
\end{align}
where we indicate pullbacks to the boundary of $\Sigma_3$ with the $\partial$ symbol. Notice that ${\bf X}_{{\bf P}_{\perp}}$ and ${\bf X}_{{\bf P}}$ have boundary terms.
The tangency of the vector field $X$ to the primary constraint submanifold implies
\begin{align*}
&0=\bm{\imath}_X  \ed\left( {\bf C}_{\perp} ( \cdot)\right)={\bf X}_{{\bf P}_{\perp}}\left(\cdot \right)=-2\int_{\partial\Sigma_3}\imath_{\partial}^*(\cdot{}){\mathrm{d}}^{\partial}Q^{\partial}\,,\\
&0=\bm{\imath}_X  \ed\left( {\bf C} (\cdot)\right)={\bf X}_{{\bf P}}\left(\cdot\right)-2\int_{\Sigma_3} \cdot \wedge{\mathrm{d}}X_Q=2\int_{\partial\Sigma_3}\imath_{\partial}^*(\cdot{})\wedge\left(\Lambda-{\mathrm{d}} {Q}_{\perp}\right)^{\partial}\,,
\end{align*}
therefore, the tangency condition of the Hamiltonian vector gives a new constraint at the boundary and fixes the Dirac multiplier $\Lambda$ \emph{only} at the boundary:
\begin{align} \label{secoPG}
\tilde{C}&:= {\mathrm{d}}^{\partial}Q^{\partial}=0\,,\qquad\qquad \Lambda^{\partial}\phantom{:}= {\mathrm{d}}^{\partial} {Q}_{\perp}^{\partial}\,.
\end{align}
Finally, we have the tangency condition
\begin{equation*}
0=\bm{\imath}_X \ed \tilde{C}= {\mathrm{d}}^{\partial} X_{Q}^{\partial}=\mathrm{d}^\partial \Lambda^\partial=\mathrm{d}^\partial \left({\mathrm{d}}^{\partial} {Q}_{\perp}^{\partial} \right)= 0\,,
\end{equation*}
therefore, there are no further constraints, and the algorithm stops.
The constraint submanifold can be described now as
\begin{eqnarray}
&&\mathcal{C}_{Pg}:=\left\{\left(Q_{\perp}, Q, {\bf P}_{\perp}, {\bf P} \right) \in T^*\mathcal{Q}:{\bf P}_{\perp}(\cdot)=0,\,  {\bf P}(\cdot)- 2\int_{\Sigma_3} \cdot \wedge{\mathrm{d}}Q=0,{\mathrm{d}}^{\partial}Q^{\partial}=0\right\}\,,
\end{eqnarray}
and the Dirac multiplier $\Lambda$ must satisfy the condition $\Lambda^{\partial}= {\mathrm{d}}^{\partial} {Q}_{\perp}^{\partial}$. The Hamiltonian vector field \eqref{NHVFPG1} reduces to
\begin{align}\label{NHVFPGF}
\begin{alignedat}{3}
X_{Q_{\perp}}   &=  \Lambda_{\perp}, \qquad\qquad\qquad& {\bf X}_{\mathbf{P}_{\perp}}   \left(\cdot\right)&=0, \\
X_{Q}           &= \Lambda\,, & {\bf X}_{\mathbf{P}}  \left(\cdot\right)&= 2\int_{\Sigma_3} \cdot  \wedge{\mathrm{d}}\Lambda\,.
\end{alignedat}
\end{align}

\medskip

\noindent \textbf{Remarks}
\begin{enumerate}\setcounter{enumi}{\value{aux}}
\item The dynamics of the pullbacks of $Q_\perp$ and $Q$ to the boundary $\partial\Sigma_3$ is determined by the pullbacks of $X_{Q_{\perp}}$ and $X_{Q}$ to the boundary\footnote{Notice that we are not pulling back a vector field but, rather, some components of the vector field are differential forms (which can be legally pulled back).}. By doing this, we get evolution equations \emph{exactly equivalent} to those found for the Chern-Simons model [as can be immediately seen by looking at the Hamiltonian vector field \eqref{HVFCSG}]. The gauge symmetry in the bulk tells us that $Q$ is completely arbitrary (although it must match a flat connection at $\partial \Sigma_3$ because ${\mathrm{d}}^{\partial}Q^{\partial}=0$ and $\Lambda^\partial={\mathrm{d}}^{\partial} {Q}_{\perp}^{\partial}$).

\item In this case, the Dirac multiplier $\Lambda$ is determined in an interesting (and somewhat unusual way) as only its boundary values are fixed. This is similar to the phenomenon observed in the case of the scalar field with Dirichlet boundary conditions that we discussed in section \ref{Sect_Dirichlet}.

\item As we have shown, it is not necessary to rely on functional differentiability to relate the Hamiltonian formulations of Chern-Simons at the boundary and the Pontryagin model in the bulk.\setcounter{aux}{\value{enumi}}

\end{enumerate}

\section{2+1 Palatini gravity with Dirichlet boundary conditions}\label{Sect_Palatini}

The final example that we discuss is a family of actions that encompass $2+1$ dimensional general relativity subject to dynamical boundary conditions of the Dirichlet type. Let $M=\mathbb{R}\times\Sigma$ be a $3$-dimensional manifold ($\Sigma$ is a 2-dimensional manifold with boundary), $G$ a semi-simple Lie group, and $\mathfrak{g}$ its Lie algebra. Let us take the action
\[
S_P(e,A)=\int_M\left(e_i\wedge F^i-\frac{\Lambda}{6}\varepsilon^{ijk}e_i\wedge e_j\wedge e_k\right)
\]
where $F$, the curvature of the $\mathfrak{g}$-valued connection $A$, is given by
\[
F^i:=\mathrm{d}A^i+\frac{1}{2}f^i_{\ jk}A^j\wedge A^k\,.
\]
In the previous expression we use abstract Latin indices to denote tensors in $\mathfrak{g}$, in particular $e_i$ is a $\mathfrak{g}^*$-valued 1-form and $f^i_{\,\,jk}$ are the structure constants (algebra indices can be raised and lowered with the Cartan-Killing metric). Also, $\varepsilon^{ijk}$ is a fixed, completely antisymmetric tensor in $\mathfrak{g}^*$ satisfying the condition $\varepsilon^{i[jk}f{^{l]}}_{im}=0$ \cite{Romano}. The field equations are now
\[F^i=\frac{\Lambda}{2}\varepsilon^{ijk}e_j\wedge e_k\,,\qquad \qquad \mathcal{D}e_i:=\mathrm{d}e_i+f_{ij}\,^{k}A^j\wedge e_k=0\,,\]
with the boundary conditions $e_i^{\partial}=0\,,$ which imply that the metric at the boundary is degenerate. We perform now the $2+1$ decomposition by writing
\begin{align*}
e_i&=e_{\perp i } {\rm d} t + e_i^\top\,,\\
A^i&=A^i_{\perp } {\rm d} t + A^{\top i}\,.
\end{align*}
In the following, we will use the simplified notation ${\mathrm{q}}=(e_{\perp i}, e_i,q^i_{\perp},q^i)$, with $q^i_{\perp}:=A^i_{\perp}$, $q^i:= A^{\top i}$, and $e_i:=e^\top_i$ (then $F^{\top i} = {\rm{d}}^\top q^i + \frac{1}{2} f^{i}_{\ jk} q^j \wedge q^k$ and $\mathcal{D}^{\top} z^i= {\rm{d}}^\top z^i + f^{i}{}_{jk} q^j \wedge z^k$). A straightforward computation gives now the Lagrangian
\[\begin{array}{cccc}
L: &  T\mathcal{Q}:= T\left( \Omega^0(\Sigma)\times \Omega^1(\Sigma)\times \Omega^0(\Sigma)\times \Omega^1(\Sigma) \right)  & \longrightarrow & \R \\
 &          \mathrm{v}=(e_{\perp}^i, e^i,q^i_{\perp},q^i; v_{e_\perp}^i,v_{e}^i, v_{q_\perp}^i,v_q^i)  & \longmapsto & L(\mathrm{v} )
\end{array} \]

\begin{eqnarray}
L(\mathrm{v})= \int_{\Sigma}  \left( v_q^i \wedge e_i +e_i \wedge \mathcal{D} q^i_{\perp} +e_{\perp i}\left( F^i-\frac{\Lambda}{2} \varepsilon^{ijk} e_j \wedge e_k \right)\right).
\end{eqnarray}
In the preceding expression, we have dropped the $\top$ superscript altogether so we write $\mathcal{D}$ instead of $\mathcal{D}^\top$ and, also, $\rm{d}$ replaces $\rm{d}^{\top}$.

If we write $\mathrm{v}:=(e_{\perp}^i, e^i,q^i_{\perp},q^i; v_{e_\perp}^i,v_{e}^i, v_{q_\perp}^i,v_q^i)$, $\mathrm{w}:=(e_{\perp}^i, e^i,q^i_{\perp},q^i;w_{e_\perp}^i,w_{e}^i, w_{q_\perp}^i,w_q^i)$ with $\mathrm{v}, \mathrm{w} \in T_{\mathrm{q}}\mathcal{Q}$, the fiber derivative is
\[
\begin{array}{l}
\displaystyle \left\langle F\!L \left(\mathrm{v}\right)|\mathrm{w}\right\rangle=\int_{\Sigma} w_q^i \wedge e_i
\end{array}
\quad \longrightarrow
\begin{array}{ll}
\displaystyle {\bf p}_{\perp i}(\cdot) := 0\,, \quad&
\displaystyle {\bf p}_{i}(\cdot)         := 0\,,\\[1.2ex]
\displaystyle {\bf P}_{\perp i}(\cdot)  := 0\,, & \displaystyle {\bf P}_{i}(\cdot):= \displaystyle \int_{\Sigma} \cdot \wedge e_i \, .
\end{array}
\]
The primary constraints are now
\begin{align}\label{conspHP}
\begin{alignedat}{3}
{\bf c}_{\perp i}(\cdot)&:= {\bf p}_{\perp i}(\cdot) = 0\,,\qquad\qquad & {\bf c}_i(\cdot)&:= {\bf p}_i(\cdot) = 0\,, \\
{\bf C}_{\perp i}(\cdot)&:= {\bf P}_{\perp i}(\cdot) = 0\,,& {\bf C}_i(\cdot)&:=  \displaystyle {\bf P}_i(\cdot)-\int_{\Sigma} \cdot \wedge e_i=0\,,
\end{alignedat}
\end{align}
and the Hamiltonian
\begin{equation}\label{Hqs}
H=-\int_{\Sigma}\left( e_i \wedge \mathcal{D} q^i_{\perp} +e_{\perp i}\left( F^i-\frac{\Lambda}{2} \varepsilon^{ijk} e_j \wedge e_k \right) \right)\,.
\end{equation}
Denoting ${ \bf p}:= \left({\bf p}_{\perp i}, {\bf p}_i, {\bf P}_{\perp i}, {\bf P}_i \right)$ we write vector fields $X \in \mathfrak{X}(T^*\mathcal{Q})$ as:
\begin{equation}
X=\Big(\left(\mathrm{q},{\bf p}\right),\left(X^i_{e_{\perp}},X^i_{e},X^i_{q_{\perp}},X^i_{q},{\bf X}_{{\bf p}_{\perp} i},{\bf X}_{{\bf p} i},{\bf X}_{{\bf P}_{\perp} i},{\bf X}_{{\bf P} i}\right)\Big)\,.
\end{equation}

The canonical symplectic form on the cotangent bundle $T^*\mathcal{Q}$  acting on pairs of vectors fields is now
\begin{align}
\begin{split}
\Omega(X,Y)&= {\bf Y}_{{\bf p}_{\perp} i}\!\left( X^i_{e_{\perp}}\right)-{\bf X}_{{\bf p}_{\perp} i}\!\left( Y^i_{e_{\perp}}\right)+{\bf Y}_{{\bf p} i}\!\left( X^i_{e}\right) - {\bf X}_{{\bf p} i}\!\left( Y^i_{e}\right)\\
&\phantom{=}+{\bf Y}_{{\bf P}_{\perp} i}\!\left( X^i_{q_{\perp}}\right)-{\bf X}_{{\bf P}_{\perp} i}\!\left( Y^i_{q_{\perp}}\right)+{\bf Y}_{{\bf P}i}\!\left( X^i_{q}\right) - {\bf X}_{{\bf P} i}\!\left( Y^i_{q}\right)\,.
\end{split}\end{align}
By following the same steps as in the preceding examples, we find that in this case the (consistent) dynamics takes place in the phase space submanifold
\[
\mathcal{C}_P:=\left\{  \left( \mathrm{q}\,, {\bf p}\right)\in T^*\mathcal{Q} :\begin{array}{llll}  {\bf p}_{\perp i}(\cdot)=0\,,&  {\bf p}_i(\cdot)=0\,,&{\bf P}_{\perp i} (\cdot)=0\,,&  {\bf P}_i(\cdot)-\displaystyle \int_{\Sigma} (\cdot) \wedge e_i = 0\,,\\
\mathcal{D} e_i=0\,, &e^{\partial}_{\perp i}=0 \,, &{e^{\partial}_i}=0\,, &\displaystyle F^{i}-\frac{\Lambda}{2} \varepsilon^{ijk} e_j \wedge e_k =0 \end{array}  \right\}\,,
\]
with $ {\lambda^i_{\perp}}^{\partial}=0$ and the Hamiltonian vector field is
\begin{align*}
  X^i_{e_{\perp}} &=  \lambda^i_{\perp}\, , & {\bf X}_{{\bf p}_{\perp} i}   \left(\cdot\right) &=0 \,, \\
 X^i_{e}       &= \mathcal{D} e^i_{\perp}- f^i{}_{jk}q^j_{\perp} e^k  \,,
                                          & {\bf X}_{{\bf p} i}  \left(\cdot\right)&= 0\,, \\
 X^i_{q_{\perp}}  &=  \mu^i_{\perp},    & {\bf X}_{{\bf P}_{\perp} i}  \left(\cdot\right)&= 0\,, \\
 X^i_{q}         &= \mathcal{D} q^i_{\perp}+\Lambda \varepsilon^{ijk} e_{\perp j} e_k \,,
                                           &  {\bf X}_{{\bf P} i}  \left(\cdot\right) &=\int_{\Sigma} \left( \cdot \right)\wedge X_{e i}\,.\end{align*}
It is straightforward to write down the integral curves for this vector field and get the gauge transformations for this model:
\begin{align}\label{NGSq}
\begin{alignedat}{3}
\delta e^i_{\perp} &=\rho^i_{\perp }\,, & \delta {\bf p}_{\perp i}  \left( \cdot  \right)  &=  0\,,\\
\delta e^i&= \mathcal{D}\rho^i - f^i{}_{jk} \tau^j e^k \,,\qquad\qquad\qquad  &\delta {\bf p}_i \left( \cdot \right)  &= 0 \,,\\
\delta q^i_{\perp} &=\tau_{\perp}^i\,, &\delta {\bf P}_{\perp i}  \left( \cdot  \right)  &=  0\,, \\
\delta q^i &= \mathcal{D}\tau^{i} +\Lambda \varepsilon^{ijk} \rho_j e_k \,,  &\delta {\bf P}_i \left( \cdot \right)  &= \int_{\Sigma} \left( \cdot \right)\wedge\big(\mathcal{D} \rho_i - f_{ijk}  \tau^j e^k \big)\,,
\end{alignedat}
\end{align}
where $\rho^i_{\perp}$, $\rho^i$, $\tau^i$, and $\tau_{\perp}^i$ are arbitrary independent functions. From the configuration variables---using the pull-back to the boundary---we get the boundary gauge symmetry
\begin{align}\label{NGSBq}
\begin{alignedat}{3}
(\delta e^i_{\perp })^{\partial}&=0\,, \qquad\qquad\qquad\qquad&(\delta q^i_{\perp})^{\partial}&= (\tau_{\perp}^{i})^{\partial}\,,\\
(\delta e^i)^{\partial}&=0\,, &(\delta q^i)^{\partial}&=\left(\mathcal{D} \tau^{i} \right)^\partial\,,
\end{alignedat}
\end{align}
where we have used ${\rho^i_{\perp}}^{\partial}=0$, ${\rho^i}^{\partial}=0$, (a consequence of  ${\lambda^i_{\perp}}^{\partial}=0$, ${e^i_{\perp}}^{\partial}=0$), and ${e^i}^{\partial}=0$.

\medskip

\noindent \textbf{Remarks}
\begin{enumerate}\setcounter{enumi}{\value{aux}}
\item The main difference between this model and standard $2+1$ Palatini gravity is the presence of non-standard boundary conditions coming from the stationarity of the action. They involve the triads, which are forced to vanish on $\partial M$. As we hope it is clear at this point, our approach allows us to effortlessly deal with them.

\item It is important to highlight the role of the condition $\varepsilon^{i[jk}f{^{l]}}_{im}=0$. Without it, new secondary constraints would arise.

\end{enumerate}

\section{Comments and concluding remarks}

As we have shown in all the examples discussed above, by judiciously relying on geometric methods it is possible to recast the Dirac algorithm in a form suitable to derive the Hamiltonian formulation of field theories defined on spatial regions with boundaries. Although the examples that we have discussed provide all the necessary information to implement the proposed method, they cannot perforce cover in an exhaustive way all the conceivable situations. In this regard, we would like to mention the possibility of finding models where constraints appear as a consequence of the need to match the values of the Lagrange multipliers (or some components of them) in the bulk and in the boundary. An interesting example where this happens is discussed in \cite{nos}.

The success and ease of use of the approach that we have followed in the paper can be traced back to the fact that \textit{direct} computations of Poisson brackets are avoided. In practice, there is no need to worry about the use of arguably formal objects, such as ``functional derivatives'', as customarily employed in the context of field theories. Problems with formal expressions for Poisson brackets are often detected when they fail to satisfy the Jacobi identity (see, for instance, \cite{Solovev1,Solovev2,Bering} for the specific case of theories with boundaries and \cite{ACZ} for a similar issue in the context of the holonomy-flux algebra in loop quantum gravity). We would like to make some comments about this.

The computation of the Poisson bracket in $T^*\mathcal{Q}$ of two (Fr\'echet) differentiable functions $f$ and $g$ amounts to finding the Hamiltonian vector fields $X_f$, $X_g$ satisfying
\begin{equation}\label{symplectic_PB}
  \bm{\imath}_{X_f}\Omega=\bm{\ed}f\,, \quad\bm{\imath}_{X_g}\Omega=\bm{\ed}g\,,
\end{equation}
and calculating
\begin{equation}\label{PBdef}
\{f,g\}:=\Omega(X_f,X_g)
\end{equation}
with the help of the canonical symplectic form $\Omega$. The Jacobi identity is then a direct and unavoidable consequence of the closedness of $\Omega$, i.e.\ if the Poisson brackets are defined according to the previous prescription (for which some requirements must be met by $f$ and $g$, in particular the obvious one of being differentiable in the mathematical sense so that $\bm{\ed}f$ and $\bm{\ed}g$ exist), then \emph{the Jacobi identity must hold}. The way to compute Poisson brackets is not to blindly use the usual expressions (modified in the case of field theories by introducing ``functional derivatives'' instead of the partial derivatives of mechanics) and, eventually, correct them by the addition/removal of boundary terms to enforce the Jacobi identity. The right way to proceed is to solve \eqref{symplectic_PB} and plug the result in \eqref{PBdef}. Although we have managed to avoid their use, Poisson brackets are, of course, very important (for instance to quantize physical models) so it is necessary to have reliable ways to compute them.

We would like to offer some thoughts on the issue of ``functional differentiability'' and its role in the present setting. There are several mathematical notions of differentiability according to the properties (in particular the topology) of the function spaces where they are defined. A very useful one is Fr\'echet differentiability in Banach spaces (that, in finite dimensional problems, reduces to the standard notion). This type of differentiability is a strong and useful regularity requirement, often necessary as a basic consistency condition for physical models. From this perspective, it is rather unfortunate to use the words ``functional differentiability'' to refer to functionals such that their variations can be written as
\[
\delta f=\int\!\!\frac{\delta f}{\delta\phi(x)}\,\delta\phi(x)\,\mathrm{d}^n x\,,
\]
with smooth $\delta f/\delta\phi(x)$ and no boundary terms. There is obviously nothing wrong with this definition but, in our opinion, it is misleading to suggest that these are the only acceptable functionals or that those involving boundary contributions are always singular. In fact, it is actually very easy to come up with examples which can be rigorously proved to be Fr\'echet differentiable\footnote{Of course this requires the explicit description of the functional spaces where they are defined.} but do not have the previous form, in particular linear and continuous maps. While in some particular instances restricting oneself to boundary-less (Regge-Teitelboim) functionals may be perfectly justified, in our opinion, this requirement is often too strong (see \cite{Solovev2} and footnote 18 in \cite[p.391]{Park}).

Although we have avoided explicit details about functional spaces, it is important to emphasize their relevance to reach a complete understanding of the Hamiltonian description of field theories with boundaries. In particular, we want to insist on the need to have precise information about the fibers of the cotangent space $T^*\mathcal{Q}$. Finally, we also want to highlight the fact that, for the purpose of quantization, what really matters is to have suitable representations of a Poisson algebra of appropriate observables from which physical predictions can be derived. In this context, functional analytic issues such as the ones commented above may conceivably be sidestepped.

\bigskip

\section*{Acknowledgments}

This work has been supported by the Spanish Ministerio de Ciencia Innovaci\'on y Universidades-Agencia Estatal de Investigaci\'on/FIS2017-84440-C2-2-P grant. Bogar D\'{\i}az is supported by the CONACYT  (M\'exico) postdoctoral research fellowship N${^{\underline{\rm{o}}}}\,$371778. Juan Margalef-Bentabol is supported by 2017SGR932 AGAUR/Generalitat de Catalunya, MTM2015-69135-P/FEDER, MTM2015-65715-P and the ERC Starting Grant 335079. He is also supported in part by the Eberly Research Funds of Penn State, by the NSF grant PHY-1806356, and by the Urania Stott fund of Pittsburgh foundation UN2017-92945.

\end{document}